\documentclass[reprint, superscriptaddress, amsmath,amssymb, aps]{revtex4-2}

\usepackage{graphicx}% Include figure files
\usepackage{dcolumn}% Align table columns on decimal point
\usepackage{bm}% bold math
\usepackage{xspace}
\usepackage{xcolor}
\usepackage{textcomp}

\begin{document}
%\linenumbers

\preprint{APS/123-QED}

\title{Modeling bacterial flow field with regularized singularities}% Force line breaks with \\

\author{Yaochen Yang}
\affiliation{Institute of Theoretical Physics, Chinese Academy of Sciences, Beijing 100190, China}
\affiliation{School of Physical Sciences, University of Chinese Academy of Sciences, 19A Yuquan Road, Beijing 100049, China}

\author{Daiki Matsunaga}
\email{daiki.matsunaga.es@osaka-u.ac.jp}
\affiliation{Graduate School of Engineering Science, The University of Osaka, Toyonaka 5608531, Japan}

\author{Da Wei}
\email{weida@iphy.ac.cn}
\affiliation{Beijing National Laboratory for Condensed Matter Physics, Institute of Physics, Chinese Academy of Sciences, Beijing 100190, China}

\author{Fanlong Meng}
\email{fanlong.meng@itp.ac.cn}
\affiliation{Institute of Theoretical Physics, Chinese Academy of Sciences, Beijing 100190, China}
\affiliation{School of Physical Sciences, University of Chinese Academy of Sciences, 19A Yuquan Road, Beijing 100049, China}
\affiliation{Wenzhou Institute, University of Chinese Academy of Sciences, Wenzhou, Zhejiang 325000, China}

\date{\today}

\begin{abstract}
The flow field generated by a swimming bacterium serves as a fundamental building block for understanding hydrodynamic interactions between bacteria. Although the flow field generated by a force dipole (stresslet) well captures the fluid motion in the far field limit, the stresslet description does not work in the near-field limit, which can be important in microswimmer interactions.
Here we propose the model combining an anisotropically regularized stresslet with an isotropically regularized source dipole, and it nicely reproduces the flow field around a swimming bacterium, which is validated by the experimental measurements of the flow field around \textit{E. coli} and our boundary-element-method simulations of a helical microswimmer, in both cases of the free space and the confined space with a no-slip wall.
This work provides a practical tool for obtaining the flow field of the bacterium, and can be utilised to study the collective responses of bacteria in dense suspensions.
\end{abstract}

\maketitle

%\tableofcontents

During the swimming process, a bacterium can generate a flow field in the surrounding fluid, which mediates its interactions with passive particles, other bacteria, etc.~\cite{lauga2016,purcell1977,berke2008,koch2011,ning2023}. To formulate this flow field, continuous efforts have been dedicated for decades. In the far field limit, the flow has been well-resolved and described as that generated by a force dipole, i.e.,  stresslet~\cite{lauga2009,batchelor1970,pedley1992}. However, when approaching the cell body of the bacterium, the above stresslet description would fail~\cite{drescher2010,drescher2011,xu2021}, primarily due to the finite size effect of the cell body \cite{wei2025,cheng2022}, and this renders the difficulties in discussing the flow field close to a bacterium and the hydrodynamic interactions between bacteria occurring at close distances~\cite{gautam2024, ishikawa2007, wensink2012}.

The above needs to understand the near-field hydrodynamics of bacteria, have motivated sustained theoretical attempts~\cite{cortez2001, zhao2019, ainley2008}, where the flow regularization method is taken as a promising candidate.
Regularization, a mathematical technique that replaces singular solutions in hydrodynamic computations with smooth analytical functions, prevents the flow field from diverging near the singularity points and facilitates approachable numerical computations~\cite{smith2018}.
For example, a regularized Stokeslet is usually implemented by replacing the point force, specifically a Dirac delta function, with a smooth function (\textit{blob} function) of certain geometry and characteristic length scale, and it has been widely utilised for modeling ciliary and flagellar flows~\cite{cortez2012,cortez2018,smith2009,zhao2021}.
For the flow field generated by a microswimmer, the description based on regularized singularities remains underexplored. This is partly because the flow field generated by a swimming microorganism—even one as simple as a bacterium—is typically complex, owing to its irregular geometry and intricate swimming mechanism. Despite some initial but meaningful efforts, such as representing the flow field by combining a regularized stresslet and rotlet dipole~\cite{ishimoto2020}, until now, these theoretical attempts do not quantitatively capture the experimentally observed near-field flow structure, which is essential for determining bacterial hydrodynamic interaction in dense suspension.
Moreover, most of the current regularization methods use isotropic blob functions, which are inherently unable to account for the anisotropy of the bacterial body and to capture the accurate near-body flow.

In this work, we develop an anisotropic regularization scheme to resolve the flow field of a swimming bacterium. By comparing with the experiments in Ref.~\cite{drescher2011} and our Boundary Element Method (BEM) simulations, we show that the flow field near a swimming \textit{E. coli} bacterium can be accurately described as that generated by an anisotropically regularized stresslet and an isotropically regularized source dipole.
This description quantitatively captures the radial flows in both the far-field and near-field limit, and also in both cases of the free space and the confined space with a no-slip wall.

\paragraph*{Theoretical model.}

\begin{figure}[htbp!]
\includegraphics[width=0.48\textwidth]{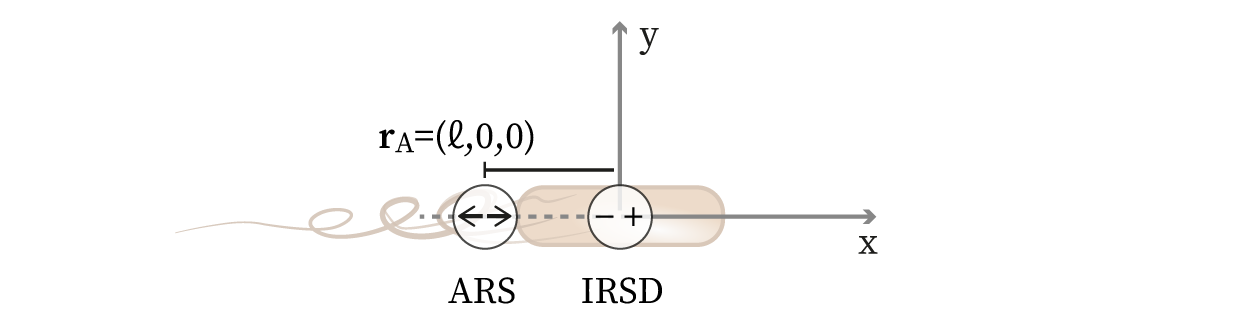}
\caption{The theoretical setup, with the ARS (anisotropically regularized stresslet) placed between the cell body and the flagellar bundle, and the IRSD (isotropically regularized source dipole) placed at the center of the cell body. The director of bacterium, $\bm{q}$, is set along the $+x$.}
\label{fig:schematic}
\end{figure}

\begin{figure*}[htbp]
\includegraphics[width=0.98\textwidth]{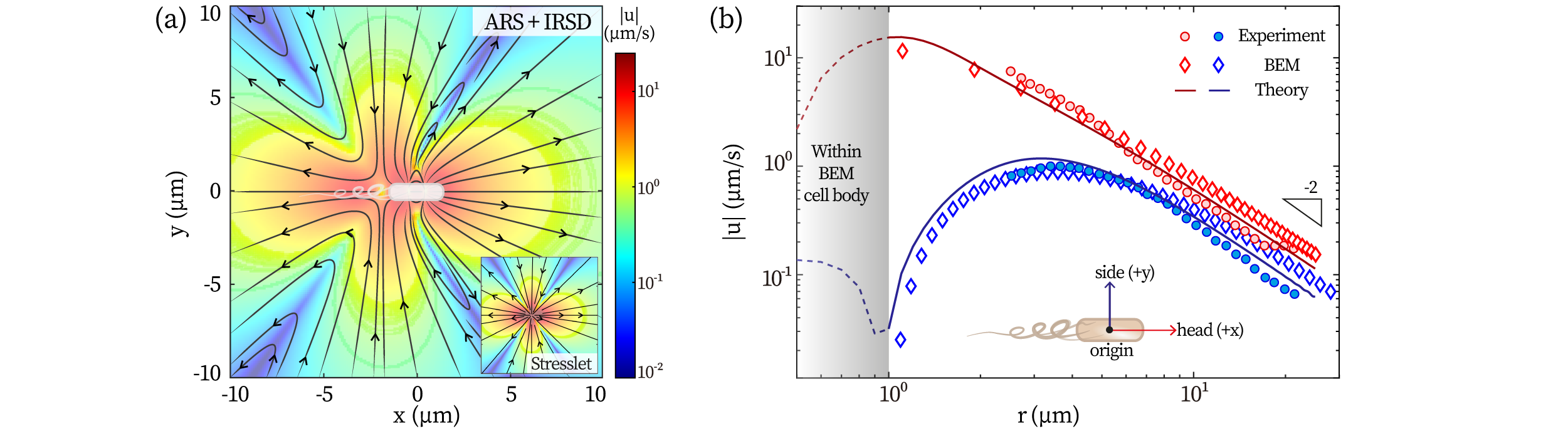}% Here is how to import EPS art
\caption{(a) \textbf{Entire flow field (ARS + ISRD) given by the model in the free space}. Colorbar indicates the magnitude of the in-plane flow velocity $|u| = \sqrt{u_x^2+u_y^2}$. Flow field of a stresslet of the same strength as the ARS is presented in the inset.  ARS: anisotropically regularized stresslet, IRSD: isotropically regularized source dipole. (b) \textbf{The head and side flow profile $|u_x(x,0,0)|$ (red) and $|u_y(0,y,0)|$ (blue).} Experimental measurements are represented by dots, numerical simulations by diamonds. Theoretical predictions inside and outside the BEM cell body surface are shown as dashed and solid lines, respectively. The fluid viscosity, $\mu = 1\times10^{-3}~\text{Pa}\cdot\text{s}$ is used in calculations throughout this work.}
\label{fig:freespace}
\end{figure*}
Consider a bacterium whose body center is located at the origin and the body axis is oriented along $+x$, as shown in Fig.~\ref{fig:schematic}, and then the flow field generated by the bacterium can be modelled as a superposition of the flow fields generated by an anisotropically regularized stresslet (ARS) and an isotropically regularized source dipole (IRSD). Note the anisotropy introduced in ARS can help to characterize the flow field induced by non-spherical shape of the swimmer.

The flow field generated by the ARS, $\bm{u}^{\mathrm{ARS}}(\bm{r})$, is constructed as:
\begin{equation}
    \bm{u}^{\mathrm{ARS}}(\bm{r}) = -\bm{D}:\nabla G_{\varepsilon,\bm{q}}(\bm{r}),
\end{equation}
where $\bm{D} = D\cdot\bm{q}\bm{q}$ denotes stresslet, and $G_{\varepsilon,\bm{q}}(\bm{r})$ denotes the Green function relating the flow field with a regularized Stokeslet, which can be obtained by solving the Stokes equation, $\nabla p - \mu\nabla^{2}\bm{u} = \bm{F} f_{\varepsilon,\bm{q}}(\bm{r})$ with $p$ as the pressure, $\mu$ the viscosity, $\bm{F}$ the force, and $f_{\varepsilon,\bm{q}}(\bm{r})$ representing the anisotropic Gaussian blob~\cite{zhao2019}.
Specifically, the anisotropic Gaussian blob function, $f_{\varepsilon,\bm{q}}(\bm{r})$, is taken as:

\begin{equation}
f_{\varepsilon,\bm{q}}(\bm{r}) = \frac{1}{\pi\sqrt{\pi}\varepsilon_\parallel \varepsilon_\perp^2}
\exp\left( -\frac{r_\parallel^2}{\varepsilon_\parallel^2} - \frac{r_\perp^2}{\varepsilon_\perp^2} \right),
\end{equation}
where the distance $r_\parallel = |\bm{q} \cdot \bm{r}|$ and $r_\perp = |\bm{r} - r_\parallel \bm{q}|$, and 
$\varepsilon_\parallel$ and $\varepsilon_\perp$ are two regularization parameters along and perpendicular to the body axis oriented in the direction $\bm{q}$, respectively. For a regularized stresslet (located at the origin), the regularization parameters generally correspond to the distance at which the flow velocities $u_x(x,0,0)$ and $u_y (0,y,0)$ reach their extrema along $x$ and $y$ axis, respectively (see more details in Supplementary Material).

Meanwhile, the flow field generated by the isotropically regularized source dipole (IRSD), $\bm{u}^{\mathrm{IRSD}}(\bm{r})$, is constructed from a regularized source monopole. The flow field by a source monopole can be expressed in the form of the gradient of a potential function: $\nabla \psi_\lambda (\bm{r})$~\cite{pozrikidis2009}, with the potential function $\psi_\lambda$ determined by the blob $f_{\lambda}(\bm{r})$, as
$\psi_\lambda(\bm{r})= -\frac{1}{4\pi}\int\frac{1}{|\bm{r}-\bm{r}'|}f_{\lambda}(\bm{r}')\ \mathrm{d}\bm{r}'$.
By taking the Gaussian blob $f_\lambda(\bm{r}) = \frac{1}{\pi\sqrt{\pi}\lambda^3}e^{-r^2/\lambda^2}$, we can obtain the analytical form of the potential function,
\begin{equation}
		\psi_\lambda(\bm{r}) = -\frac{1}{4\pi r} \text{erf}(\frac{r}{\lambda}).
\end{equation}
Then the flow field generated by IRSD is given by,
\begin{equation}
    \bm{u}^{\mathrm{IRSD}}(\bm{r}) =  - \bm{d}\cdot\nabla\nabla\psi_\lambda(\bm{r}),
\end{equation}
where $\bm{d} = d\cdot \bm{q}$ denotes the source dipole.

Then, the entire flow field in the laboratory frame, is given by the superposition of the above two flow field generated by the ARS and IRSD. For simplicity, we set $\bm{q}$ as $+x$ direction. Since the center of the cell body is located at the origin and flagellar bundle extends to $x<0$, ARS center is placed at $\bm{r}_A = (-l,0,0)$, as shown in Fig.~\ref{fig:schematic}. Because IRSD enforces no-slip condition at the two ends of the long axis of the cell body, which is symmetric regarding the origin, its center coincides with the geometric center of the cell body. Therefore, the entire flow field is given by,

\begin{equation}
    \bm{u}(\bm{r}) = \bm{u}^{\text{ARS}}(\bm{r}-\bm{r}_A) + \bm{u}^{\text{IRSD}}(\bm{r}).
\end{equation}
Note that, since the regularization does not alter the far-field behavior of singularity solutions, the strength of both the stresslet $\bm{D}$ and the source dipole $\bm{d}$ are defined in the same fashion as those in the case without regularization. For readers familiar with the squirmer model~\cite{ishikawa2024review,ishimoto2013, graaf2016,lintuvuori2016}, a higher-order multipole summation is usually needed for accurately capturing the flow field near the cell body, which is computationally heavy.

\begin{figure*}[htbp]
\includegraphics[width=0.98\textwidth]{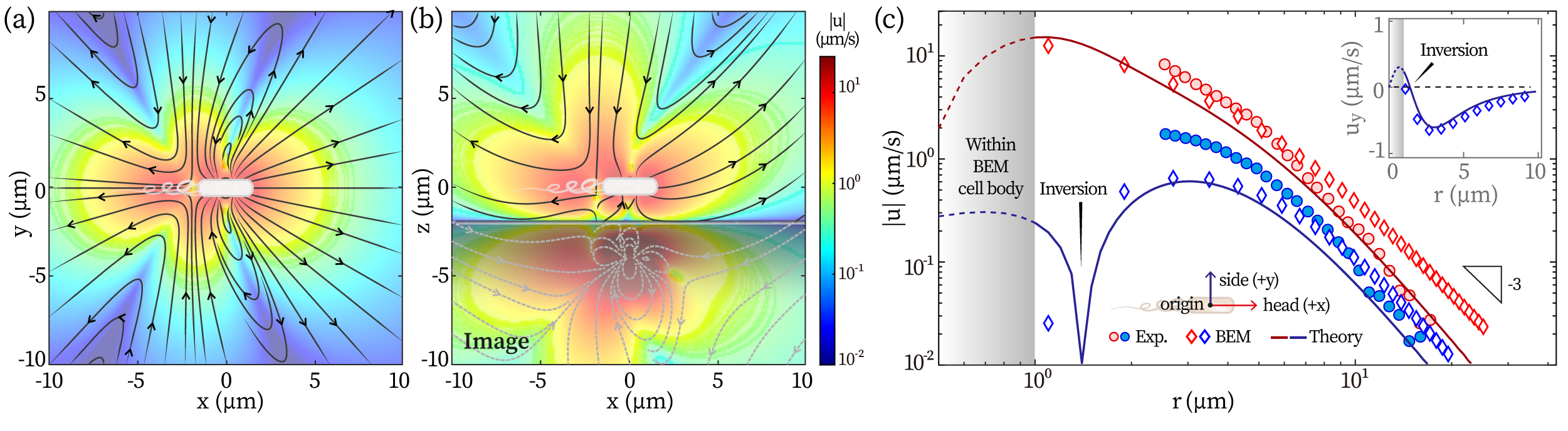}
\caption{(a-b) \textbf{Entire flow field given by the model in presence of a no-slip wall.} Colorbar shows the magnitude of the in-plane flow velocity: (a) $|u| = \sqrt{u_x^2+u_y^2}$ and (b) $|u| = \sqrt{u_x^2+u_z^2}$. (c) \textbf{The head and side flow profile $|u_x(x,0,0)|$ and $|u_y(0,y,0)|$.} Experimental measurements are represented by dots, numerical simulations by diamonds. Theoretical predictions inside and outside the BEM cell body surface are shown as dashed and solid lines, respectively. The inset shows the side flow profile $u_y(0,y,0)$ around the inversion point.}
\label{fig:nearwall}
\end{figure*}

\paragraph*{Comparisons with experiments and simulations.}
We compare the theoretical results obtained by the above model with both the experiments on \textit{E. coli}~\cite{drescher2011} and our BEM simulations.
In BEM simulations, the bacterial body is simplified as a sphere of 1~$\mu m$ radius and the flagellar bundle is represented by a helical tail of 10~$\mu m$ length in the swimming direction ($x$ direction), and one can refer to Supplementary Material for more details.
The comparisons are focused on the axial velocity along the $+x$ axis, $u_x(x,0,0)$ ($x>1~\mu \text{m}$), and the lateral velocity along the $+y$ axis, $u_y(0,y,0)$ ($y>1~\mu \text{m}$), where $1$~$\mu m$ is chosen due to the fact that the size of the cell body in BEM simulations is $1$~$\mu m$.
These velocities are referred to as the \emph{head} and \emph{side} velocity in following discussions, respectively.
We shall discuss the flow fields generated by the bacterium in both the free space and the confined space with a no-slip wall.

In the free space, both the experimental and BEM results are accurately captured by the theoretical predictions using the parameter set $\{D = 1\ \text{pN}\cdot\mu \text{m},l = 1.6\ \mu \text{m},\varepsilon_\parallel = 1~\mu \text{m},\varepsilon_\perp = 1.36\ \mu \text{m} ,d = 140\ \mu \text{m}^4/\text{s} ,\lambda = 0.7\ \mu \text{m}\}$. 
Note that the parameter $D$ and $l$ have explicit physical correspondence, which are the stresslet strength~\cite{drescher2011,hu2015} and half body length~\cite{linek2016,chattopadhyay2006,wei2024} of \textit{E. coli}, respectively. Thus, the primary fitting parameters in our system are the regularization lengths of the stresslet, i.e., $\{ \varepsilon_\parallel,\varepsilon_\perp\}$. Since $\varepsilon_\parallel$ and $\varepsilon_\perp$ are highly coupled, we fix $\varepsilon_\parallel=1~\mu$m and fit only $\varepsilon_\perp$ (correlated with the body geometry, see Supplementary Material) throughout this work. In the secondary fitting step, the IRSD is introduced to better capture the surface flow velocity in the $x$-direction and to “finetune” the region within $\sim$1 body length from the cell, and the parameters, $\{d,\lambda\}$, are fitted independently. This sequential fitting procedure avoids overfitting of the system.

The pattern of the flow field in the $xy$-plane, calculated by the model, is provided in Fig.~\ref{fig:freespace}(a).
In the far field limit, the flow velocity decays in the form of $\propto r^{-2}$ as shown in in Fig.~\ref{fig:freespace}(b), same as that by a singular stresslet. This is due to the fast decay of the flow field given by IRSD, which is in the form of $\propto r^{-3}$, and the effect of IRSD then becomes negligble in the far field limit.
In the near field limit, the model captures two main features observed in experiments and BEM simulations:
(a) the head flow profile, $u_x(x,0,0)$, decays monotonically and the swimming speed is close to that obtained in BEM simulations, 11.8~$\mu m$/s; (b) the side flow profile, $u_y(0,y,0)$, has an initial increase followed by a decrease in the form of $\propto r^{-2}$, and the flow speed approaches zero at the surface $y = 1~\mu \text{m}$, due to the flow inversion induced by the stresslet center offset $l$.

Then we generalise the comparisons to the case in the confined space with a no-slip wall, where the wall confinement is important in changing the flow field~\cite{meng2021,cheng2024}.
Theoretically, we need to calculate the flow field by considering the wall effect, by
considering the contributions of image systems~\cite{blake1971,cortez2015}:
\begin{eqnarray}
		\bm{u}^{\text{ARS}}_{\text{wall}}(\bm{r}) &=& \bm{u}^{\text{ARS}} + \bm{u}^{\text{ARS}}_{\text{img}}\\
		\bm{u}^{\text{IRSD}}_{\text{wall}}(\bm{r}) &=& \bm{u}^{\text{IRSD}} + \bm{u}^{\text{IRSD}}_{\text{img}}.
\end{eqnarray}
The image system of ARS is derived by anisotropically regularizing the Blake tensor and the image system of IRSD is derived based on the image system of an isotropically regularized source monopole (see Supplementary Material for more details).
To compute the flow field near the wall in BEM simulations, the bacterium is placed above a no-slip wall with the distance $2~\mu \text{m}$. All other settings are kept identical to those in the case of the free space.

The flow field parallel to the surface ($xy$-plane) and perpendicular to the surface ($xz$-plane) are shown in Figs.~\ref{fig:nearwall} (a) and (b), respectively. The flow profiles along the head and side directions are shown in Fig.~\ref{fig:nearwall}(c).
The flow field in the near field limit behaves similarly as in the free space, whereas the velocity decaying in the far field limit follows the form of $\propto r^{-3}$ due to the no-slip wall.
Note that both the theoretical results and BEM simulations show that the extremum of the side flow profile is lower than that in the free space, whereas the experimental measurements show the opposite trend: the extremum of the side flow profile being higher in the presence of the wall.
This discrepancy can originate from a spiral flow field, which is usually described by a pair of opposite rotlets \cite{ishimoto2020,Lopez2014,lauga2016}. Note that when comparing $u_y$ in Fig.~\ref{fig:nearwall}, we have removed the contribution of the rotlet dipole by taking $u_y\left(0,y,0\right)=\left[u_y^\prime\left(0,y,0\right)-u_y^\prime\left(0,-y,0\right)\right]/2$ where $u_y^\prime$ denotes the raw flow field in BEM (see Supplementary Material).

\paragraph*{Summary.}
We develop an analytical model for describing the flow field generated by a swimming bacterium, in both the far field and the near field limit, based on the combination of regularized singularities.
The theoretical results are validated by comparisons with experiments and our BEM simulations.
The anisotropy introduced in the model nicely captures the geometry of a bacterium, and thus the flow field.
The approach provides a complementary view to the multipole expansion of microswimmer flow field~\cite{chwang1975,lighthill1975} and can boost further investigations in flow fields generated by other microorganisms~\cite{drescher2010,wei2019,Wei2021}.
Moreover, the flow field predicted by the model can be utilised to describe the hydrodynamic interactions (forces and torques) between bacteria in dense suspensions, which is important in determining the collective dynamics of bacteria, e.g., bacterial turbulence.

\begin{acknowledgments}
We thank Knut Drescher for the informational input on interpreting the experimental data, and Ramin Golestanian for discussion.
F.M. acknowledges supports by the National Natural Science Foundation of China (NSFC) (Grant No. 12275332 and 12447101), Max Planck Society (Max Planck Partner Group), Wenzhou Institute (Grant No. WIUCASQD2023009), and Beijing National Laboratory for Condensed Matter Physics (Grant No. 2023BNLCMPKF005) and UCAS Xiaomi Youth Fellowship. D.W. acknowledges supports by NSFC (Grant No. 12574238). D.M. acknowledges supports by the Japan Society for the Promotion of Science KAKENHI (Grant 21H05879, 23K22673, 23H04418, and 23K26040) and the Japan Science and Technology Agency PRESTO (Grant No. JP-MJPR21OA).
\end{acknowledgments}

\bibliography{reference}% Produces the bibliography via BibTeX.

\end{document}

\setlength{\parindent}{0pt} %

\title{Supplementary Material of \\
 Modeling bacterial flow field with regularized singularities}% Force line breaks with \\

\author{Yaochen Yang}
\affiliation{Institute of Theoretical Physics, Chinese Academy of Sciences, Beijing 100190, China}
\affiliation{School of Physical Sciences, University of Chinese Academy of Sciences, 19A Yuquan Road, Beijing 100049, China}

\author{Daiki Matsunaga}
\email{daiki.matsunaga.es@osaka-u.ac.jp}
\affiliation{Graduate School of Engineering Science, Osaka University, Toyonaka 5608531, Japan}

\author{Da Wei}
\email{weida@iphy.ac.cn}
\affiliation{Beijing National Laboratory for Condensed Matter Physics, Institute of Physics, Chinese Academy of Sciences, Beijing 100190, China}

\author{Fanlong Meng}
\email{fanlong.meng@itp.ac.cn}
\affiliation{Institute of Theoretical Physics, Chinese Academy of Sciences, Beijing 100190, China}
\affiliation{School of Physical Sciences, University of Chinese Academy of Sciences, 19A Yuquan Road, Beijing 100049, China}
\affiliation{Wenzhou Institute, University of Chinese Academy of Sciences, Wenzhou, Zhejiang 325000, China}

\maketitle

\section{Integral representation of anisotropically regularized stokeslet}

\begin{figure}[htb!]
    \centering
    \includegraphics[width=0.5\linewidth]{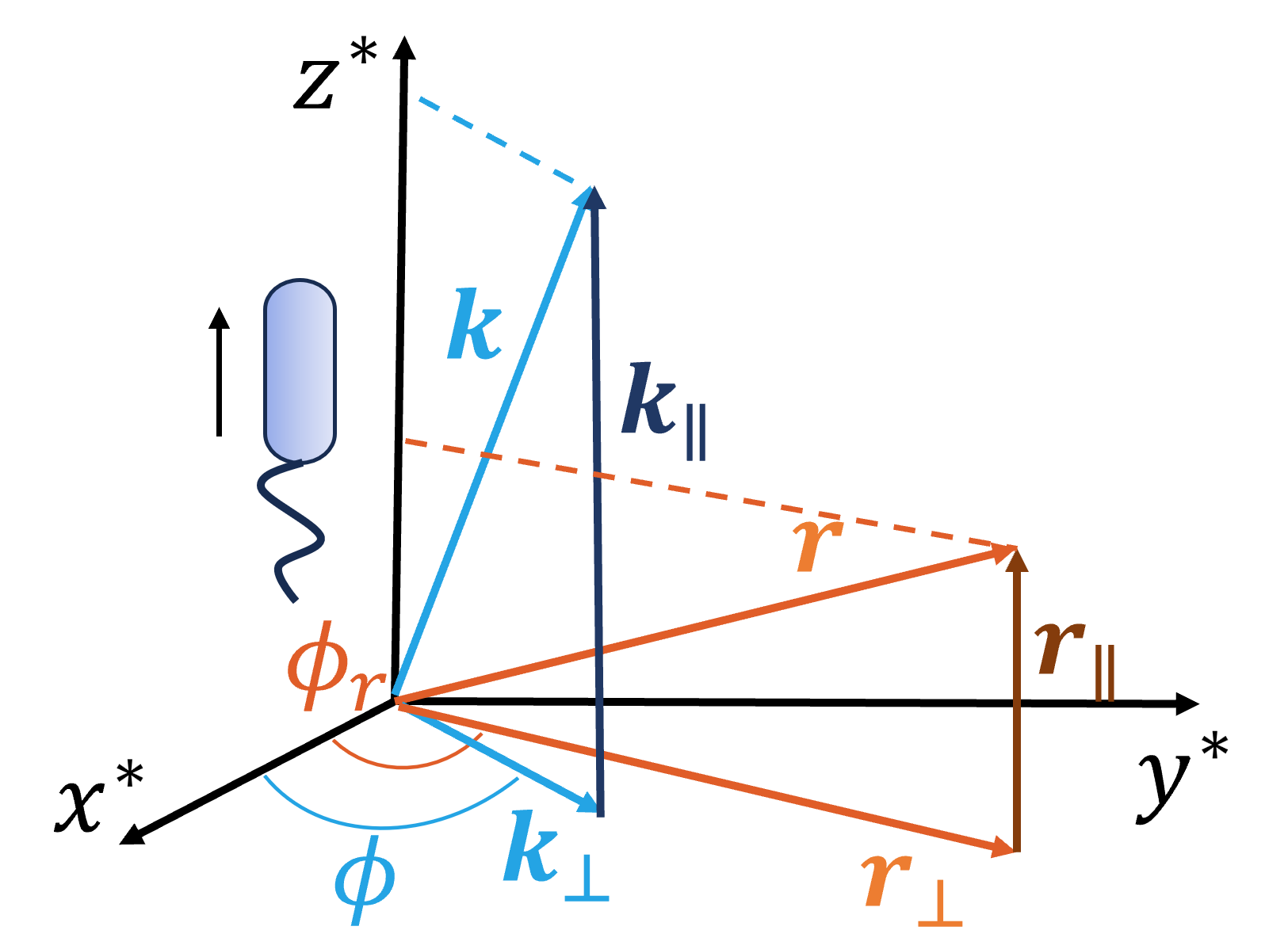}
    \caption{\textbf{Schematic drawing of the cylindrical coordinate system $(k_\perp,\phi,k_\parallel)$ in the body-fixed frame.} The bacterium swims along $z^*$-direction.}
    \label{fig:schematic integral}
\end{figure}

The Green tensor of anisotropically regularized stokeslet in the coordinate space is represented by the inverse Fourier transform of $\tilde{G}$:
\begin{equation}
    \bm{G}_{\varepsilon,\bm{q}} = \frac{1}{(2\pi)^3}\int\frac{1}{\mu k^2}(\bm{I}-\frac{\bm{k}\bm{k}}{k^2})\tilde{f}_{\varepsilon,\bm{q}}(\bm{k})e^{i\bm{k}\cdot\bm{r}}\ \mathrm{d}\bm{r}.
\end{equation}
Considering the symmetry of the system, the inverse transformation is conveniently computed in the bacterium's body-fixed frame shown in Fig.~\ref{fig:schematic integral}, as $\tilde{f}_{\varepsilon,\bm{q}}(\bm{k})$ will maintain a concise analytical form:
\begin{equation}
    \tilde{f}_{\varepsilon,\bm{q}}(\bm{k}) = e^{-(\varepsilon_\parallel^2k_\parallel^2+\varepsilon_\perp^2k_\perp^2)/4}.
\end{equation}
We perform the integration in a cylindrical coordinate system $(k_\perp, \phi, k_\parallel)$ with the $\bm{q}$ (bacterial orientation axis) aligned along $+k_\parallel$. 
\begin{equation}
    \bm{G}_{\varepsilon,\bm{q}} = \frac{1}{(2\pi)^3}\int_0^\infty\!\!\!\!\int_0^{2\pi}\!\!\!\!\int_{-\infty}^{+\infty}\!\!\frac{1}{\mu k^2}(\bm{I}-\frac{\bm{k}\bm{k}}{k^2})\tilde{f}_{\varepsilon,\bm{q}}(\bm{k})e^{i\bm{k}\cdot\bm{r}}k_\perp\ \mathrm{d}k_\perp \mathrm{d}\phi \mathrm{d}k_{\parallel}, 
\end{equation}
where $\bm{k}\cdot\bm{r} = k_\perp r_\perp\cos{(\phi-\phi_r)}+k_\parallel r_\parallel$ and $\phi_r$ is the azimuthal angle of $\bm{r}$. The integral is simplified by firstly integrating the azimuthal angle $\phi$, defining intermediate integral $A_{iso}$ and $A_{ij}$:
\begin{equation}
    \begin{dcases}
        A_{iso} = \int_0^{2\pi}e^{ik_\perp r_\perp\cos{(\phi-\phi_r)}}\ \mathrm{d}\phi,\\
        A_{ij} = \int_0^{2\pi}\frac{k_ik_j}{k^2}e^{ik_\perp r_\perp\cos{(\phi-\phi_r)}}\ \mathrm{d}\phi.\\
    \end{dcases}
\end{equation}

\begin{table}[h]
\centering
\caption{\textbf{Integral representation of $A_{iso}$ and $A_{ij}$.}}
\label{tab:integral-array}
\begin{tabular}{cccc}
    \hline\hline
    $A_{ij}$ & \text{integrand} & \text{result}  \\ \hline
    $A_{iso}$ & $\displaystyle e^{ik_\perp r_\perp\cos{(\phi-\phi_r)}}$ & 
    $\displaystyle 2\pi J_0(k_\perp r_\perp)$ \\
    $A_{xx}$ & $\displaystyle \cos^2\!\!{\phi}e^{ik_\perp r_\perp\cos{(\phi-\phi_r)}}$ & 
    $\displaystyle \pi[J_0(k_\perp r_\perp)-J_2(k_\perp r_\perp)]\cos{2\phi_r}$ \\
    $A_{yy}$ & $\displaystyle \cos^2\!\!{\phi}e^{ik_\perp r_\perp\cos{(\phi-\phi_r)}}$ & 
    $\displaystyle \pi[J_0(k_\perp r_\perp)+J_2(k_\perp r_\perp)]\cos{2\phi_r}$ \\
    $A_{zz}$ & $\frac{k_\parallel^2}{k^2}e^{ik_\perp r_\perp\cos{(\phi-\phi_r)}}$ &
    $\displaystyle 2\pi \frac{k_\parallel^2}{k^2} J_0(k_\perp r_\perp)$\\
    $A_{xy},A_{yx}$ & $\displaystyle\cos{\phi}\sin{\phi}e^{ik_\perp r_\perp\cos{(\phi-\phi_r)}}$ &  $\displaystyle -\pi J_2(k_\perp r_\perp)\sin{2\phi_r}$ \\
    $A_{xz},A_{zx}$ & $\displaystyle \frac{k_\parallel}{k}\cos{\phi}e^{ik_\perp r_\perp\cos{(\phi-\phi_r)}}$ & $\displaystyle 2\pi iJ_1(k_\perp r_\perp)\cos{\phi_r}$\\
    $A_{yz},A_{zy}$ & $\displaystyle \frac{k_\parallel}{k}\sin{\phi}e^{ik_\perp r_\perp\cos{(\phi-\phi_r)}}$ & $\displaystyle 2\pi iJ_1(k_\perp r_\perp)\sin{\phi_r}$\\
    \hline\hline
\end{tabular}
\end{table}
one has:
\begin{equation}
    G_{ij}^{\varepsilon,\bm{q}} = \frac{1}{(2\pi)^3}\int_0^\infty\!\!\!\!\int_{-\infty}^{+\infty}\!\!\frac{1}{\mu k^2}(A_{iso}\delta_{ij}-A_{ij})\tilde{f}_{\varepsilon,\bm{q}}(\bm{k})e^{ik_\parallel r_\parallel}k_\perp \ \mathrm{d}k_\perp \mathrm{d}k_\parallel.
\end{equation}
The simplified integral exhibits favorable numerical convergence under the transformation $k_\perp\to\frac{\xi}{1-\xi},~|k_\parallel|\to\frac{\chi}{1-\chi},~ \xi\in(0,1),~\chi\in(0,1)$. Finally, the results are rotated back to the laboratory frame using the rotation tensor $\bm{\mathscr{R}}$:
\begin{equation}
    \bm{G}_{\varepsilon,\bm{q}}^{lab-frame} = \bm{\mathscr{R}}\cdot\bm{G}_{\varepsilon,\bm{q}}^{body-frame}\cdot\bm{\mathscr{R}}^{-1},
\end{equation}
as $\bm{\mathscr{R}}\cdot\hat{\bm{z}}^* = \hat{\bm{x}}$.

\clearpage

\section{Image systems}
\subsection{Image system for ARS}
The standard tensor form of a flow field given by the Blake tensor reads \cite{blake1971}:
\begin{equation}
    \bm{u}^{\text{Blake}} = \frac{1}{8\pi\mu}\Big[(\frac{\bm{I}}{r}+\frac{\bm{r}\bm{r}}{r^3}-\frac{\bm{I}}{R}-\frac{\bm{R}\bm{R}}{R^3})\cdot\bm{F}+2(\bm{F}_\parallel-\bm{F}_\perp)\cdot\nabla_{\bm{R}}(\frac{\bm{R}\bm{h}}{R^3}-\frac{\bm{I}}{R}-\frac{\bm{R}\bm{R}}{R^3})\cdot\bm{h}\Big].
\end{equation}
Generalized to anisotropically regularized case:
\begin{equation}
    \bm{u}^{\text{AR-Blake}}(\bm{r}) = \int\bm{u}^{\text{Blake}}(\bm{r}')f_{\varepsilon,\bm{q}}(\bm{r}-\bm{r}')\ \mathrm{d}(\bm{r}-\bm{r}'),
\end{equation}
where definition of $\bm{r},\bm{r}',\bm{R}$ is shown in the schematic draw. Flow field of an anisotropically regularized stokeslet near a wall reads:
\begin{eqnarray}
    \bm{u}^{\text{AR-Blake}}(\bm{r}) &=&\int\bm{G}(\bm{r}')\cdot\bm{F}f^\varepsilon_{\bm{q}}(\bm{r}-\bm{r}')\ \mathrm{d}(\bm{r}-\bm{r}')\nonumber\\
    &-&\int\bm{G}(\bm{R}')\cdot\bm{F}f^\varepsilon_{\bm{q}}(\bm{R}-\bm{R}')\ \mathrm{d}(\bm{R}-\bm{R}')\nonumber\\
    &-&\int2(\bm{F}_\parallel-\bm{F}_\perp)\cdot\nabla_{\bm{R}'}\bm{G}(\bm{R}')\cdot\bm{h}'f^\varepsilon_{\bm{q}}(\bm{R}-\bm{R}')\ \mathrm{d}(\bm{R}-\bm{R}')\nonumber\\
    &+&\int2(\bm{F}_\parallel-\bm{F}_\perp)\cdot\nabla_{\bm{R}'}\bm{U}(\bm{R}')
    h'^2f^\varepsilon_{\bm{q}}(\bm{R}-\bm{R}')\ \mathrm{d}(\bm{R}-\bm{R}'),
\end{eqnarray}
where $\bm{G}(\bm{r}) = \frac{1}{8\pi\mu}(\frac{\bm{I}}{r}+\frac{\bm{r}\bm{r}}{r^3}),~ \bm{U}(\bm{r}) = \frac{1}{8\pi\mu}\frac{\bm{r}}{r^3}$. The first two terms are both from the anisotropically regularized stokeslet $\bm{G}_{\varepsilon,\bm{q}}$, the last two terms need detailed calculation. Here we introduce an approximation $\bm{h}'\to h$ to avoid complex integration. This approximation may introduce a small boundary residual near the center of the ARS, but its impact on the overall flow field is negligible.

\begin{eqnarray}
    \int \nabla_{\bm{R}'}\bm{G}(\bm{R}')\cdot\bm{h}f^\varepsilon_{\bm{q}}(\bm{R}-\bm{R}')\ \mathrm{d}(\bm{R}-\bm{R}')  = \nabla_{\bm{R}}\bm{G}_{\varepsilon,\bm{q}}(\bm{R})\cdot\bm{h}
\end{eqnarray}

\begin{eqnarray}
    \int \nabla_{\bm{R}'}\bm{U}(\bm{R}'){h}'^2f^\varepsilon_{\bm{q}}(\bm{R}-\bm{R}')\ \mathrm{d}(\bm{R}-\bm{R}')= h^2\nabla_{R}\bm{U}(\bm{R})
\end{eqnarray}

The image system for $G_{\varepsilon,\bm{q}}(\bm{r})$ is then:
\begin{equation}
    \bm{G}^{\text{AR}-Stokes}_{\text{img}}(\bm{r}) \approx - \bm{G}_{\varepsilon,\bm{q}}(\bm{R}) + 2\nabla_{\bm{R}}\Big[\bm{G}_{\varepsilon,\bm{q}}(\bm{R})\cdot\bm{h}-h^2\bm{U}(\bm{R})\Big]^T\cdot(\frac{2\bm{hh}}{h^2}-\bm{I}).
\end{equation}
The image system for ARS is obtained as a 3-order tensor:
\begin{equation}
    \bm{\Delta}(\bm{r})^{\text{ARS}}_{\text{img}} = - \nabla\bm{G}^{\text{AR}-Stokes}_{\text{img}}(\bm{r}),
\end{equation}
note that $\bm{h}$ should be treated as a variable in the derivative. The corresponding image flow reads:
\begin{equation}
    \bm{u}^{\text{ARS}}_{\text{img}} = \bm{D}:\bm{\Delta}(\bm{r})^{\text{ARS}}_{\text{img}}.
\end{equation}
\begin{figure}[htb!]
    \centering
    \includegraphics[width=0.5\linewidth]{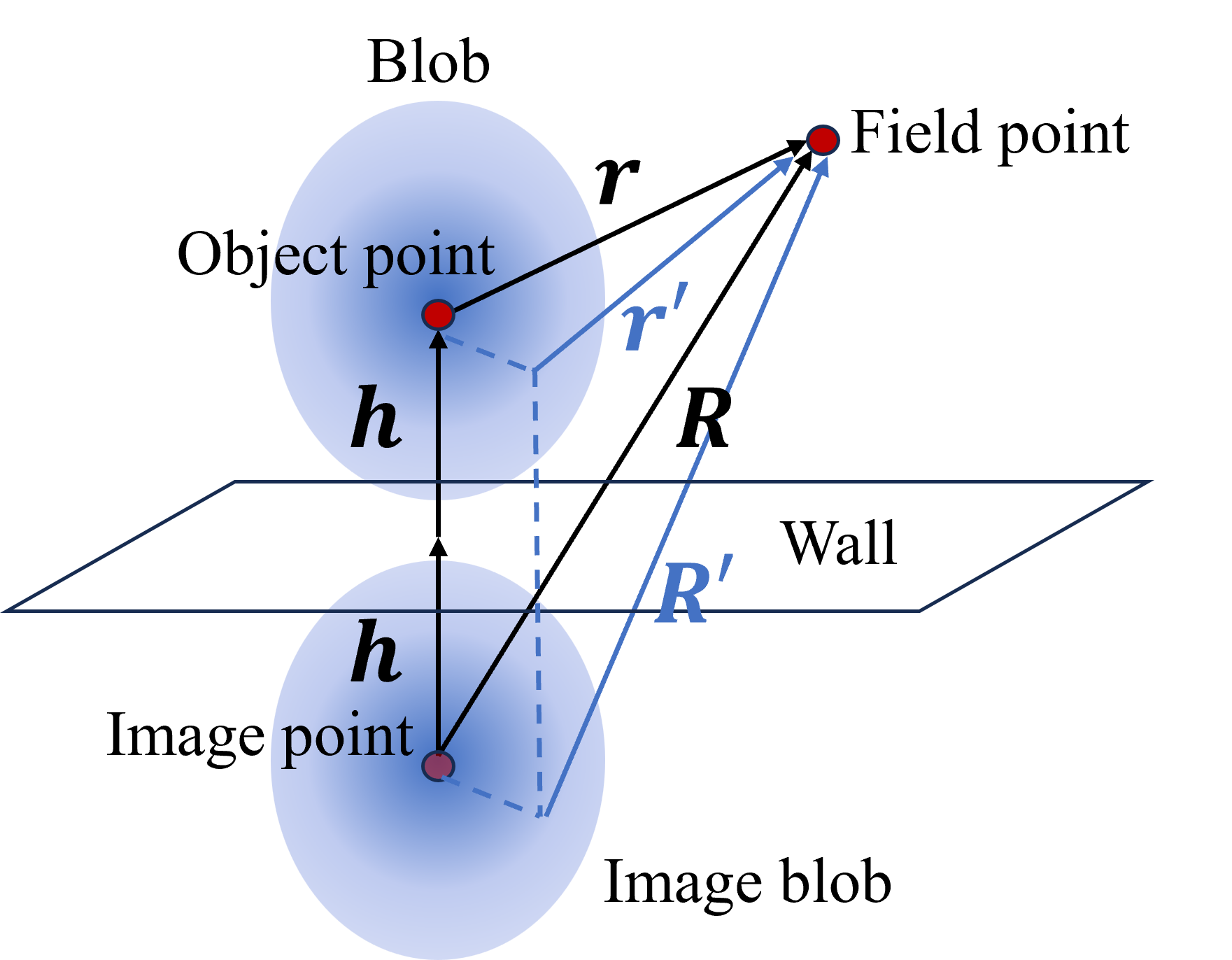}
    \caption{\textbf{Schematic drawing of the image systems}.}
    \label{fig:schematic image}
\end{figure}
\subsection{Image system for IRSD} 
The image system of a regularized source $\nabla\psi_\lambda(\bm{r})$ is derived by Cortez \cite{cortez2015}:
\begin{eqnarray}
    \bm{\sigma}^{\text{image}}_\lambda = \nabla\psi_\lambda(\bm{r}) -4\Big\{\frac{\bm{h}\bm{h}}{h^2}:\mu\nabla\bm{G}_\lambda(\bm{R})-\frac{1}{2}\bm{h}\cdot[f_\lambda(\bm{R})\bm{I}-\nabla\nabla\psi_\lambda(\bm{R})]\Big\},
\end{eqnarray}
where $\bm{G}_\lambda$ is isotropically regularized stokeslet.
The image system for IRSD is obtained as a 2-order tensor:
\begin{equation}
    \bm{H}^{\text{IRSD}}_{\text{img}} = -\nabla\bm{\sigma}^{\text{image}}_\lambda,
\end{equation}
note that $\bm{h}$ should still be treated as a variable in the derivative. The corresponding image flow reads:
\begin{equation}
    \bm{u}^{\text{IRSD}}_{\text{img}} = \bm{d}\cdot\bm{H}^{\text{IRSD}}_{\text{img}}.
\end{equation}

\subsection{Additional validation of the near-wall case}

\begin{figure}[htb!]
    \centering
    \includegraphics[width=0.6\linewidth]{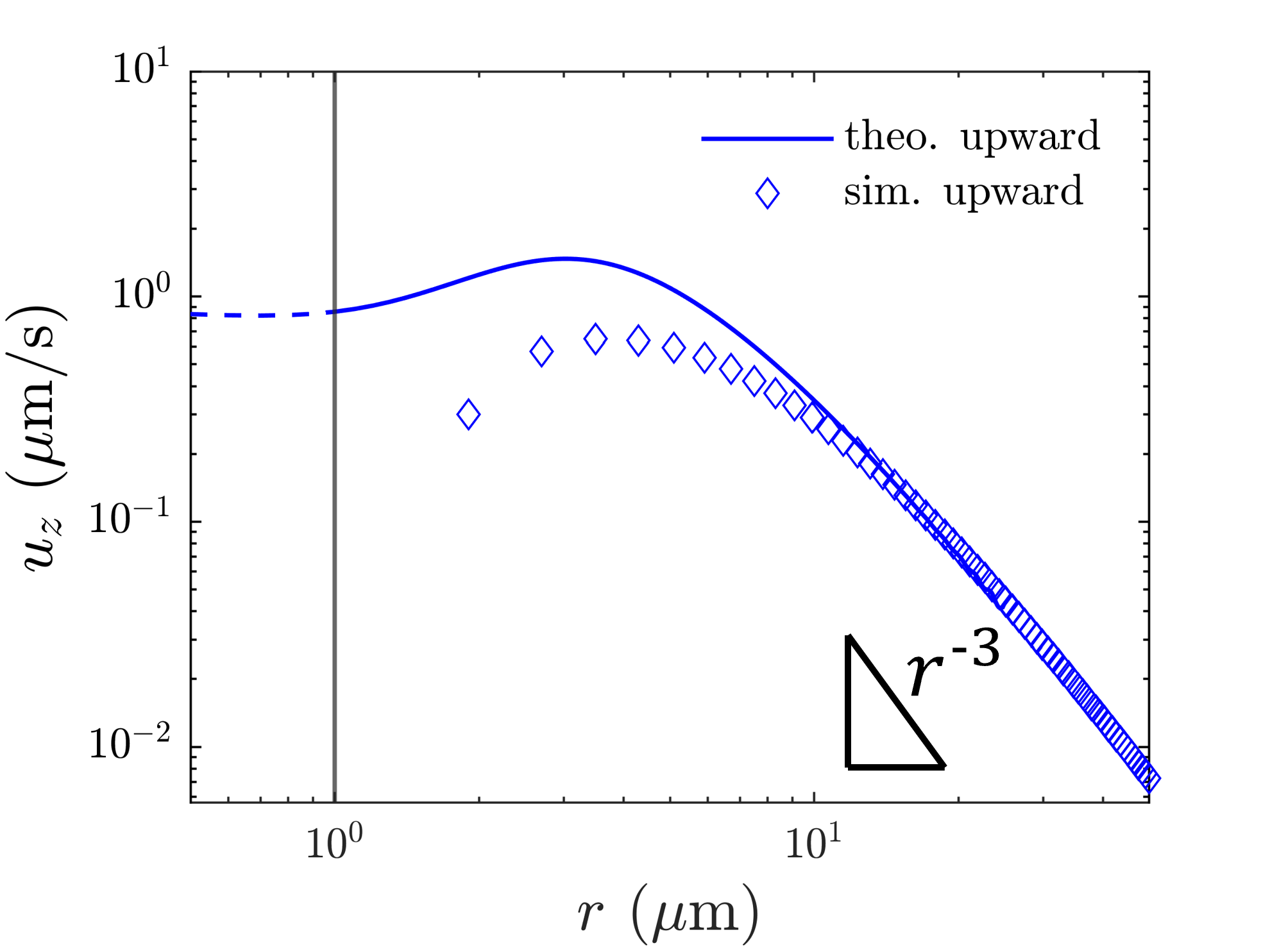}
    \caption{\textbf{Additional validation of the near-wall case.} Flow profile $u_z (0,0,r)$ for $z>0$ (upward flow) is shown, where vertical solid line shows the cell body surface at $r=1 \mu\text{m}$.}
    \label{fig:additional validation}
\end{figure}
The flow profile $u_z (0,0,r)$ for $z>0$ (upward flow) is shown in Fig.~\ref{fig:additional validation}. The deviation at small $r$ is due to the blob of the ARS overlapping with its wall-induced image, since the cell body is too close to the wall.

\clearpage
\section{BEM Simulation}
\subsection{Equations}
We use Boundary Element Method (BEM) to simulate a model swimming organism and investigate its flow field. This swimmer consists of a spherical head of radius $a$ and a helical tail of length $10a$ along $x$-direction, which rotate in opposite angular velocity.
The velocity at a given material point on the swimmer surface $\bm{r}_s$ can be expressed in terms of the translational velocity $\bm{U}$ and the rotational velocity $\bm{\Omega}$ as:
\begin{equation}
    \bm{u}(\bm{r}_s) = \bm{U} + \bm{\Omega} \times (\bm{r} - \bm{r}_0). 
\end{equation}
The flow velocity at a given position $\bm{r}$ can be obtained using the boundary integral equation:
\begin{equation}
    \bm{v}(\bm{r}) = -\frac{1}{8\pi\mu} \int_{S} \bm{G}\cdot \bm{t} \, \mathrm{d}S,
\end{equation}
where $\bm{t}$ is the traction vector.
At the swimmer surface, $\bm{u}(\bm{r}_s) = \bm{v}(\bm{r}_s)$, leading to the following equation:
\begin{equation}
    \bm{U} + \bm{\Omega} \times (\bm{r} - \bm{r}_0) + \frac{1}{8\pi\mu} \int_S \bm{G}\cdot \bm{t} \, \mathrm{d}S = \mathbf{0}.
\end{equation}
The system satisfies force-free and torque-free conditions:
\begin{eqnarray}
    \begin{dcases}
        \bm{F} = \int_{S} \bm{t} \, \mathrm{d}S = \mathbf{0}, \\
    T_{yz} = \left( \int_{S} (\bm{r} - \bm{r}_0) \times \bm{t} \ \mathrm{d}S \right)_{yz} = \mathbf{0}, \\
    T_{x}^{\mathrm{head}} = \left( \int_{S} (\bm{r} - \bm{r}_0) \times \bm{t} \ \mathrm{d}S \right)_{x}^{\mathrm{head}} = +T, \\
    T_{x}^{\mathrm{tail}} = \left( \int_{S} (\bm{r} - \bm{r}_0) \times \bm{t} \ \mathrm{d}S \right)_{x}^{\mathrm{tail}} = -T.
    \end{dcases}
\end{eqnarray}
Here $+T$ and $-T$ represent a pair of rotlet dipoles. Note that in the main text, the contribution of the rotlet dipole has been subtracted from the BEM flow field data.

\subsection{Dimensionless Forms}
The torque equation can be converted to dimensionless form:
\begin{eqnarray}
    \begin{dcases}
        T_{x}^{\mathrm{head}*} = \left( \int_{S} (\bm{r}^{*} - \bm{r}_{0}^{*}) \times \frac{a^{3} \bm{t}}{T} \, \mathrm{d}S^{*} \right)_{x}^{\mathrm{head}} = +1, \\
        T_{x}^{\mathrm{tail}*} = \left( \int_{S} (\bm{r}^{*} - \bm{r}_{0}^{*}) \times \frac{a^{3} \bm{t}}{T} \, \mathrm{d}S^{*} \right)_{x}^{\mathrm{tail}} = -1.
    \end{dcases}
\end{eqnarray}
The dimensionless traction is defined as \(\bm{t}^{*} = (a^{3} / T) \bm{t}\). The velocity can also be non-dimensionalized:
\begin{equation}
    \frac{\mu a^2}{T} \bm{U} + \frac{\mu a^3}{T} \bm{\Omega} \times (\bm{r}^* - \bm{r}_0^*) + \frac{1}{8\pi} \int_S \bm{G}^* \bm{t}^* \, \mathrm{d}S^* = \mathbf{0}.
\end{equation}
where $\bm{U}^{*} = (\mu a^{2} / T) \bm{U}$ and $\bm{\Omega}^{*} = (\mu a^{3} / T) \bm{\Omega}$. The simulation yields the following results:
\begin{equation}
    \begin{dcases}
        U_{x}^{*} &= -1.31 \times 10^{-3} \\
\Omega_{x}^{\mathrm{head}*} &= +3.91 \times 10^{-2} \\
\Omega_{x}^{\mathrm{tail}*} &= -6.98 \times 10^{-3}.
    \end{dcases}
\end{equation}
To compare with an experimental system we have $a = 1~\mu\text{m},~U= 11.8~\mu\text{m}/\text{s}$. Note that $T = \mu a^2\frac{U}{U^*} = 9~\text{pN}\cdot\mu\text{m}$ is significantly greater than that of a real bacterium \cite{Darton2007}. Real \textit{E. coli} has elongated cell body, multiple flagella, and complex flagellar structure, which endows them with higher swimming efficiency, thereby leading to this discrepancy.
\clearpage

\section{Parameter determination and justification}
We designed a step-by-step parameter determination procedure to systematically determine the model parameters. Among these parameters, $D$ and $l$ have explicit physical correspondence, which are the stresslet strength~\cite{drescher2011,hu2015} and half body length~\cite{linek2016,chattopadhyay2006,wei2024} of \textit{E. coli}, respectively. Thus, the primary fitting parameters in our system are the regularization lengths of the stresslet, i.e., $\{ \varepsilon_\parallel,\varepsilon_\perp\}$. Since $\varepsilon_\parallel$ and $\varepsilon_\perp$ are highly coupled, we fix $\varepsilon_\parallel=1~\mu$m and fit only $\varepsilon_\perp$ (correlated with the body geometry, see Supplementary Material) throughout this work. In the secondary fitting step, the IRSD is introduced to better capture the surface flow velocity in the $x$-direction and to “finetune” the region within $\sim$1 body length from the cell, and the parameters, $\{d,\lambda\}$, are fitted independently. This sequential fitting procedure avoids overfitting of the system.

\setlength{\tabcolsep}{10pt}
\begin{table}[h]
\centering
\caption{\textbf{Parameter determination procedure.} The \textbf{data} column indicates the data source or target data of fitting. The corresponding physical dimension for each parameter is listed as: $D(\text{pN}\cdot\mu \text{m}),~ d(\mu\text{m}^4/\text{s}),~\{l,~\varepsilon_{\parallel,\perp},~\lambda,~r\}(\mu \text{m})$.}
\label{tab:fitting procedure}
\begin{tabular}{cccc}
    \hline\hline
   \text{step}  & \text{parameter}& \text{method} & \text{data set} \\ \hline
    1 & $D$ & \text{experiments} & \text{\cite{drescher2011,hu2015}}\\
    2 & $l$ & \text{experiments} & \text{\cite{wei2024}}\\
    3 & $\varepsilon_\parallel$ & \text{fixed} & $1~\mu\text{m}$ \\
    4 & $\varepsilon_\perp$ & \text{fitting} & $u_y(0,y,0)$\\
    5 & $\lambda$ & \text{fitting} & $x(\text{surface})$ \\ 
    6 & $d$ & \text{fitting} & $u_x(x,0,0)$\\
    \hline\hline
\end{tabular}
\end{table}

\clearpage
\section{Validation by additional BEM}

\begin{figure}[htb!]
    \centering
    \includegraphics[width=0.9\linewidth]{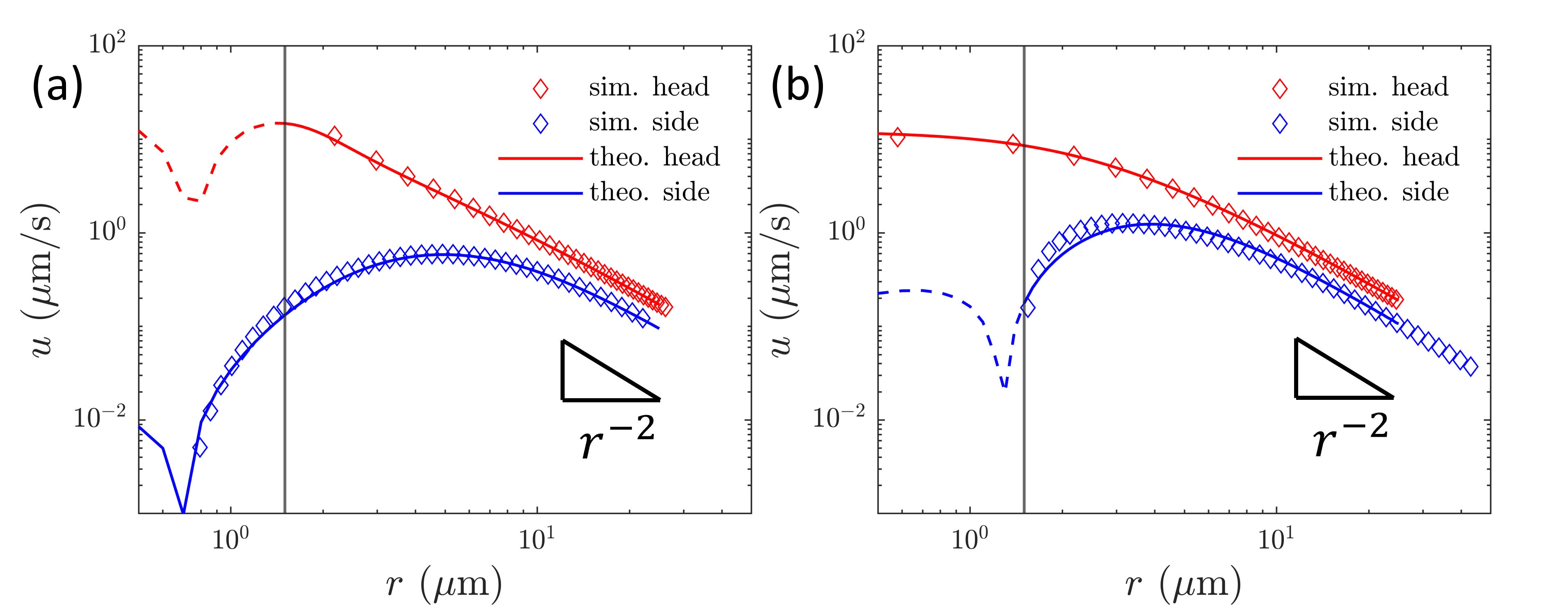}
    \caption{\textbf{Validation by additional BEM simulations of varied body shapes in free space.}.  The head and side flow profile $|u_x(x,0,0)|$ and $|u_y(0,y,0)|$ for (a) prolate cell body and (b) oblate cell body.}
    \label{fig: fitting additional BEM}
\end{figure}

In the main text, we validated the model by comparing with both experiments and BEM (sphereical body shape). Here we perform additional BEM simulations with varied cell body geometries (a prolate ellipsoid with an aspect ratio of 3:1 and an oblate one with 1:3 aspect ratio) in free space, in order to check how the body geometry affects the regularization parameters.
The model fits well to BEM simulations for different cell body shapes (see Fig.~\ref{fig: fitting additional BEM} and Tab.~\ref{tab:datasets}), with the parameter set as: $\{D = 1.65\ \text{pN}\cdot\mu \text{m},l = 3\ \mu \text{m},\varepsilon_\parallel = 1~\mu \text{m},\varepsilon_\perp = 3.04\ \mu \text{m} ,d = 400\ \mu \text{m}^4/\text{s} ,\lambda = 0.9\ \mu \text{m}\}$ for the prolate case and $\{D = 1.75\ \text{pN}\cdot\mu \text{m},l = 2.1\ \mu \text{m},\varepsilon_\parallel = 1~\mu \text{m},\varepsilon_\perp = 1.84\ \mu \text{m} ,d = 0\}$ for the oblate case.

\setlength{\tabcolsep}{10pt}
\begin{table}[h]
\centering
\caption{\textbf{Four datasets used in this work.}} 
\label{tab:datasets}
\begin{tabular}{cccccc}
    \hline\hline
     \text{dataset} & $\varepsilon_\parallel$ & $\varepsilon_\perp$ & $e$ & $x_{\max}/\varepsilon_\parallel$ &$y_{\max}/\varepsilon_\perp$\\ \hline
    \text{EXP.} + \text{BEM} & 1 & 1.36 & 0.68 & 1.6 & N.A.\\
    \text{BEM (spherical)} & 1 & 2.27 & 0.90 & 2.4 & 0.88\\
    \text{BEM (prolate)} & 1 & 3.04 & 0.94 & 3.0 & 0.59\\
    \text{BEM (oblate)} & 1 & 1.84 & 0.84 & 2.1 & 0.97\\
    \hline\hline
\end{tabular}
\end{table}

\clearpage

\section{Sensitivity analysis}
\begin{figure}[htb!]
    \centering
    \includegraphics[width=1\linewidth]{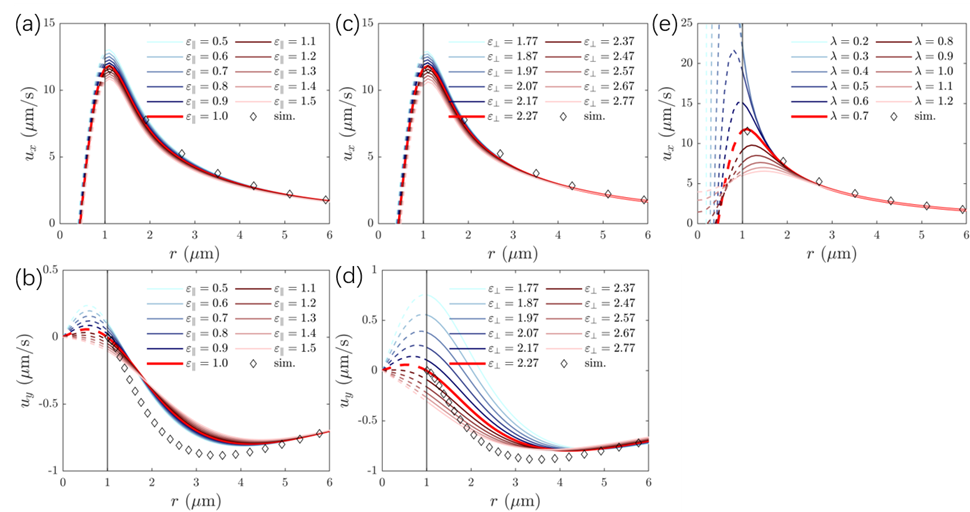}
    \caption{\textbf{Sensitivity of the regularization parameters $\varepsilon_\parallel,\varepsilon_\perp,\lambda$ ($\mu\text{m}$)}.  Panels (a) and (b) show the response of $u_x (x,0,0)$ and $u_y (0,y,0)$, respectively, to the variation of $\varepsilon_\parallel$, and (c) and (d) show the response of the same quantities to the variation of $\varepsilon_\perp$. Panel (e) shows the response of $u_x (x,0,0)$ to the variation of $\lambda$; the source dipole does not contribute to $u_y (0,y,0)$. When varying one parameter, the others are fixed to their default values ($\varepsilon_\parallel$,$\varepsilon_\perp$,$\lambda$)= (1, 2.27, 0.7). All lengths are in $\mu$m.}
    \label{fig:sensitivity}
\end{figure}

We perform sensitivity analysis of the regularization parameters $\varepsilon_\parallel,\varepsilon_\perp,\lambda$ (FIG~\ref{fig:sensitivity}). We find that $u_y$ is particularly sensitive to variations in $\varepsilon_\parallel$, while $u_x$ is insensitive to both $\varepsilon_\parallel$ and $\varepsilon_\perp$. Meanwhile, $\lambda$ significantly influences the magnitude and the extremum location of $u_x$ near the cell body surface at $x=1~\mu\text{m}$.

\clearpage
\section{Necessity of anisotropic regularization}
Anisotropic regularization captures the inherent flow anisotropy around a bacterium, which standard isotropic regularization fails to reproduce. 
As shown in experimental and BEM flow fields, this built-in anisotropy manifests through a difference between $x_{\rm max}$ and $y_{\rm max}$, the respective distance where $|u_x|$ and $|u_y|$ maximize. An isotropically regularized stresslet always gives $x_{\rm max}/y_{\rm max}=1$ and thus cannot describe such near-field behavior. Note that the above analysis is based on the origin being located at the stresslet center. $x_{\rm max}$ and $y_{\rm max}$  are proportional to $\varepsilon_\parallel$ and $\varepsilon_\perp$, respectively, with the proportionality constants influenced by the shape of the ARS blob, which is characterized by the eccentricity $e=\sqrt{1-\varepsilon_\parallel^2/\varepsilon_\perp^2}$.

\begin{figure}[htb!]
    \centering
    \includegraphics[width=0.9\linewidth]{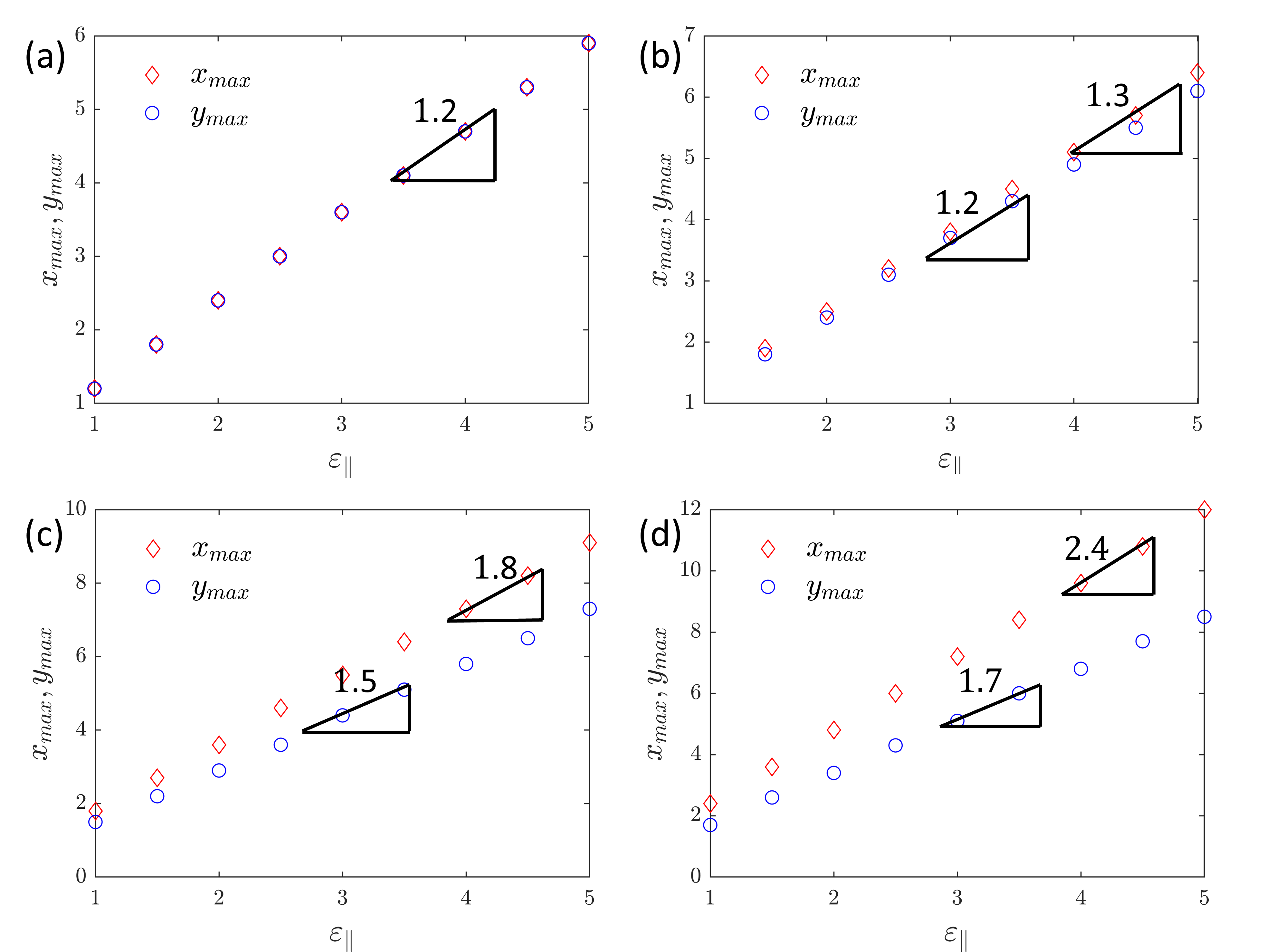}
    \caption{Relation between $x_{\rm max}$, $y_{\rm max}$ as a function $\varepsilon_\parallel$ for different eccentricity (a) $e=0$, (b) $e=0.4$, (c) $e=0.8$, and (d) $e=0.9$.}
    \label{fig:extrmum}
\end{figure}

\begin{figure}[htb!]
    \centering
    \includegraphics[width=0.9\linewidth]{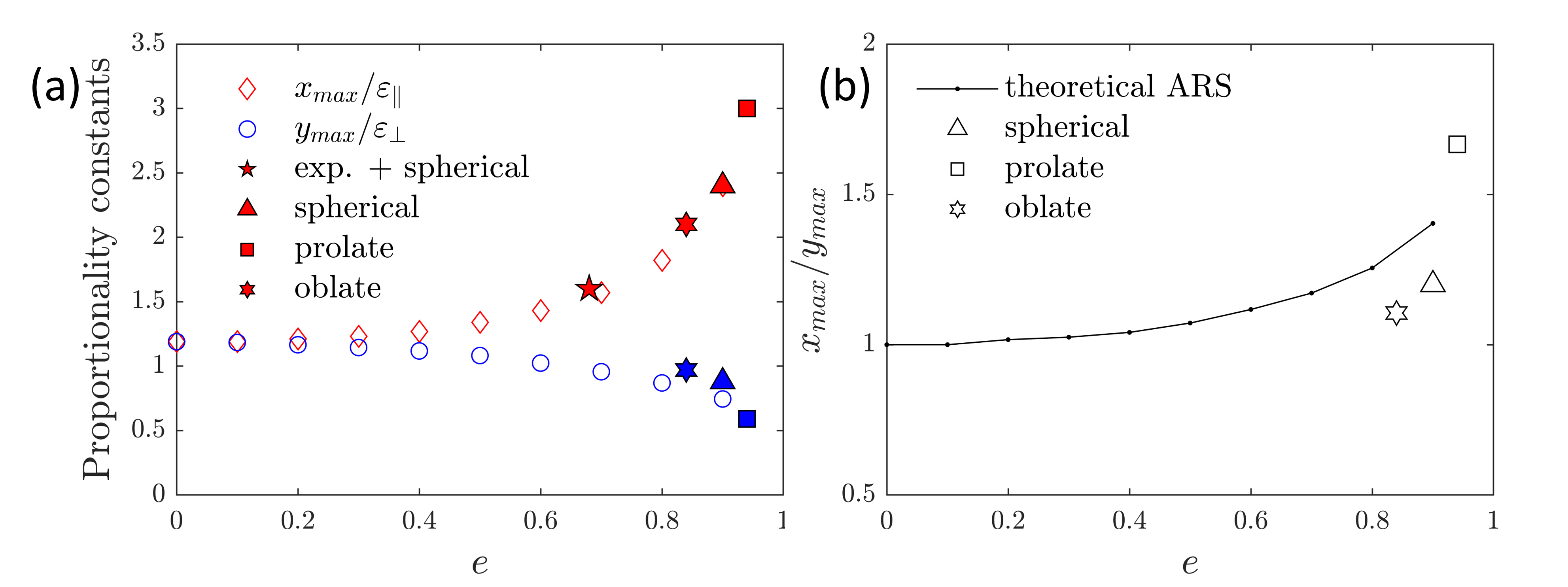}
    \caption{(a) The proportionality constants, $x_{\rm max}/\varepsilon_\parallel$ and $y_{\rm max}/\varepsilon_\perp$, and (b) $x_{\rm max}/y_{\rm max}$, as a function of the eccentricity $e$ ($\varepsilon_\parallel$ is fixed as 1~$\mu$m here).}
    \label{fig:ratio}
\end{figure}

\clearpage
\section{About rotlet dipole}
\textit{E. coli} generates a spiral flow field near the cell body and flagellar bundle, due to their opposite rotation, which is usually described by a pair of opposite rotlets (or a rotlet dipole). flow field of a rotlet with torque $+\bm{T}$ reads:
\begin{equation}
    \bm{u}^{\text{rot}}(\bm{r}) = \frac{1}{8\pi\mu}\frac{\bm{T}\times\bm{r}}{r^3}.
\end{equation}
Note that when comparing $u_y$ in the main text, we have removed the contribution of the rotlet dipole by taking $u_y\left(0,y,0\right)=\frac{1}{2}\left[u_y^\prime\left(0,y,0\right)-u_y^\prime\left(0,-y,0\right)\right]$ where $u_y^\prime$ denotes the raw flow field in BEM.

Rotlet and its dipole are also suitable for regularizaion, both isotropic and anisotropic,
\begin{figure}[htpb!]
    \centering
    \includegraphics[width=1\linewidth]{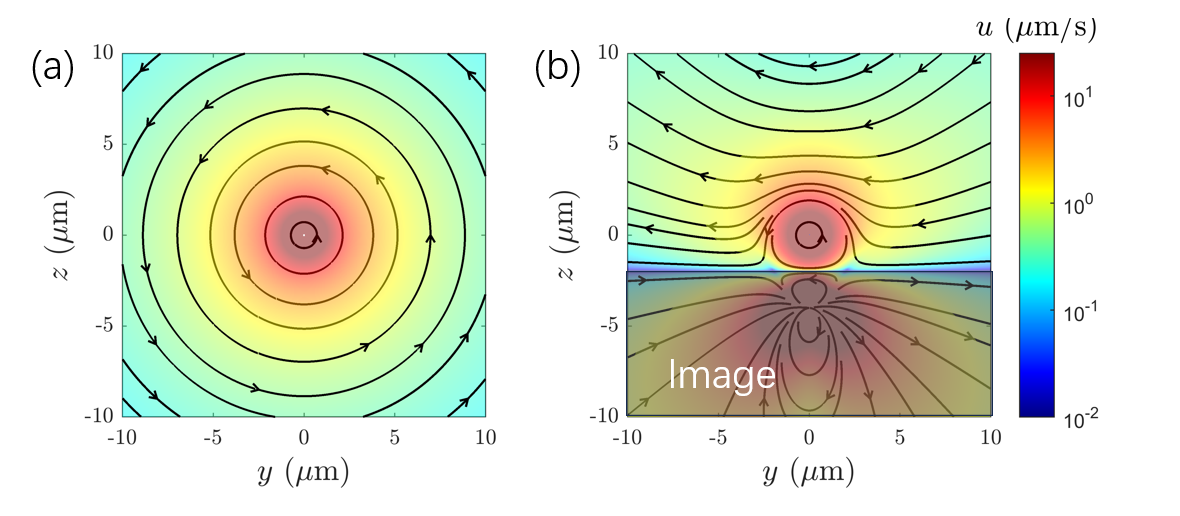}
    \caption{\textbf{Flow field in $yz$-plane of a rotlet.} (a) In free space. (b) Near a wall.}
    \label{fig: rotlet}
\end{figure}

\clearpage
\bibliography{reference}